\documentclass[onecolumn,preprintnumbers,amsmath,prc,amssymb]{revtex4}

\usepackage{graphicx}
\usepackage{dcolumn}
\usepackage{bm}

\begin{document}

\title{On bimodal size distribution of spin clusters in the one dimensional Ising model}

\author{A. I. Ivanytskyi and V. O. Chelnokov}
\affiliation{Bogolyubov Institute for Theoretical Physics of the National Academy of Sciences of Ukraine, Metrologichna str. 14$^b$, Kiev-03680, Ukraine}



\begin{abstract}
The size distribution of  geometrical spin clusters is exactly found for the one dimensional Ising model 
of finite extent. For the values of  lattice constant $\beta$ above some ``critical value" $\beta_c$ the found size 
distribution demonstrates  the non-monotonic behavior with the peak corresponding to the size of  largest available 
cluster. In other words, at high values of  lattice constant there are two ways to fill the lattice: either to form a single 
largest cluster or to create many clusters of small sizes.  This feature closely resembles the well-know  bimodal size 
distribution of clusters which is usually interpreted as a robust  signal of  the first order liquid-gas phase transition in 
finite systems. It is remarkable that the bimodal size distribution of spin  clusters  appears in the  one dimensional Ising 
model  of finite size, i.e. in the model which in thermodynamic limit  has no phase transition at all.

\vspace*{0.5cm}


\end{abstract}

\maketitle


\section{Introduction}
\label{Intro}

During last two decades the experimental studies  of phase transitions in finite and even in small systems are  inspiring  
the high interest to their rigorous theoretical  treatment  \cite{Gross,ComplMoretto,Bmodal:Chomaz01,CSMM}. One of 
the main reasons  for  such an interest is that the nuclear systems which are experimentally studied at intermediate 
\cite{JoeNat,Pichon,Bonnet} and at high  \cite{Pratt,Karsch13} collision energies have no thermodynamic limit due to the presence
of the long range Coulomb interaction and, hence, in a strict thermodynamical  sense  the phase transition in such systems 
is not defined. Therefore, the practical need to formulate the reliable experimental signals of phase transformations in the 
systems consisting of a few hundreds or thousands particles  lead researchers to   development of  the non-traditional 
statistical methods which may be suited to small systems \cite{Bmodal:Chomaz01,CSMM,Chomaz03,Bugaev05,Bmodal:Bugaev13}. 
One of such directions of research, the concept of bimodality,  is based on the old T. L. Hill idea \cite{THill:1} that  in finite 
system the interface between two pure phases ``costs" additional free energy and, hence, their coexistence is suppressed. 
The practical conclusion coming out of this idea is that the resulting distribution of the order parameter should demonstrate a 
bimodal behavior  and, hence,  each maximum or peak of the bimodal distribution has to be  associated with the pure phase 
\cite{THill:1,Bmodal:Chomaz01, Chomaz03,Bmodal:Chomaz03,Bmodal:Gulm04,Bmodal:Gulm07}. 

Although in \cite{Bmodal:Chomaz03} the authors claimed to establish the one-to-one correspondence between the bimodal 
structure of   specially  constructed  partition of some measurable quantity, known on average,  and the properties of the 
Lee-Yang zeros \cite{YangLee} of  this  partition in the complex plane  of  fugacity of this measurable quantity, 
there appeared certain doubts 
\cite{CSMM,Lopez, Bmodal:Bugaev13} about so strict relation between  the bimodal-like mass/charge distributions of nuclear 
fragments and  the liquid-gas nuclear phase transition in finite systems. In particular,  the  doubts follow from the experimental  
analysis of high momentum decay channels in nucleus-nucleus collisions \cite{Lopez} and from the exact analytical solutions  
of the simplified statistical multifragmentation model \cite{Mekjian} found   for finite \cite{Bugaev05,Bmodal:Bugaev13,CSMM} 
and  infinite \cite{Sagun14} volumes. Two explicit counterexamples  suggest that  the bimodal  mass distribution of nuclear 
fragments   appears in a finite volume analog of gaseous phase  \cite{Bmodal:Bugaev13} with positive surface tension and that 
it can also occur in the thermodynamic limit at supercritical temperatures \cite{Sagun14}, if  at such temperatures  there exists  
negative surface tension.  

Despite   the  existing  counterexamples \cite{Bmodal:Bugaev13,Sagun14}, up to now it is unclear, whether the statistical 
systems without phase transition can, in principle,  generate the bimodal  size distributions of their constituents for finite 
volumes.  In order to demonstrate that this is, indeed, the case and the statistical systems in finite volume can  generate the 
bimodal  size distributions, here we analytically calculate the size distribution of geometrical spin clusters of the simplest 
statistical model which has no phase transition, the one dimensional Ising model  \cite{Ising}. Another principal purpose of 
this work is to  further develop  connections between the spin models and cluster models.  

The cluster models  are successfully applied to study  phase transitions in simple liquids \cite{Fisher,Dillmann,Ford}, in nuclear 
matter \cite{Bondorf,CSMM,Sagun14,Mekjian,Bmodal:Bugaev13,Bugaev05} and in the quark-gluon plasma phenomenology  
\cite{Kapusta, Gorenstein,QGBSTM,QGBSTM2,QGBSTM3,CGreiner:06,CGreiner:07,Koch:09}. Their basic assumption is that the 
properties of  studied system consisting from  the elementary degrees of freedom (molecules, nucleons, partons, etc.) can be 
successfully described in terms of  physical clusters (vapor droplets, nucleus,  hadrons, etc.) formed by  any natural  number 
of the elementary degrees of freedom. Since  the cluster models proved to be a successful tool to investigate the  phase 
transition mechanisms both in finite and in infinite systems, it seems that  the development of  equivalent cluster formulation for  
the well-known  spin models  and the determination of the properties of their  physical clusters is a theoretical   task of high priority.
It is necessary to mention that some results on such a connection  were already  reported in  \cite{ComplMoretto}, in which a 
parameterization of the Fisher liquid droplet model \cite{Fisher} was successfully applied to the numerical  description of  cluster 
multiplicities  of two- and three-dimensional Ising  model. However,  below we  implement this program to the one dimensional 
Ising model and directly calculate the size distribution of the geometrical spin clusters from the corresponding partition function.

The work  is organized as follows. In the next section we explain the necessary mathematical aspects of calculating  the spin  cluster 
size distribution. Then, in Sect. 3 the derived  expression is analyzed in details and the physical origin of bimodality in the present 
model is discussed. The results of our numerical simulations  are also presented in this section. Finally, our conclusions and perspectives 
of further research are  summarized in Sect. 4.

 
\section{Cluster size distribution} 
\label{Distribution}

Here we use the geometrical definition of clusters which seems to be  the most natural for  lattice models. Within this 
framework the monomers, the dimers, the  trimers and so on are built from the neighboring spins of the same direction. 
This approach is useful  for different  lattice systems \cite{ComplMoretto,Moretto,Fortunato2000,Fortunato2001}. 
A high level of its generality allows one to study the SU(2) and SU(3) gauge lattice models in terms of the clusters  
composed from the  Polyakov loops \cite{Gattringer2010,Gattringer2011,Regensburg15, Kiev15}. Let's consider the one 
dimensional Ising model in which spins are  arranged in a linear lattice of size $N$ and they can be in two states only 
(up or down). Interaction energy of two neighboring spins is proportional to their product. Hence, the neighboring  pair 
of parallel spins contribute $-\epsilon$ to the total energy of the system, whereas the corresponding contribution of the 
neighboring  pair  of  antiparallel spins is  $\epsilon$. For a convenience we choose the following boundary conditions: two  
edge spins have only a single neighbor which is able to interact with them. 

Evidently,  in the present model there are only   two types of clusters, i.e. the clusters composed of  spins up and  of spins down.  
Hence, for a convenience hereafter  they are called the up  clusters and the down clusters and are marked with the superscripts 
$\uparrow$ and $\downarrow$, respectively.

The cluster multiplicities  are defined as  their occupancy numbers $n^\uparrow_k$ and $n^\downarrow_k$, where $k$ is the cluster
size. It is clear that size of the maximal cluster can not exceed $N$ and, hence,  only the  spin configurations obeying the condition 
$n^\uparrow_{k>N}=n^\downarrow_{k>N}=0$ can be  realized in the considered  system. The total numbers of up and down spin clusters are, 
respectively, denoted as $n^\uparrow\equiv\sum_k n^\uparrow_k$ and $n^\downarrow\equiv\sum_k n^\downarrow_k$. Hereafter, 
it is assumed that the blind index of cluster size runs over all positive integers. In what follows  the set of all occupancy numbers 
$\sigma\equiv\{n^\uparrow_k,n^\downarrow_k\}$ is called as  a microscopic configuration. Each  microscopic configuration $\sigma$ 
of the system  defines its total energy, which receives a contribution $\epsilon(1-k)$ from every  up cluster or  from  every down cluster 
of size $k$. In addition, there exist the  contacts of  neighboring clusters with  opposite spins   and their number is  
$n^\uparrow+n^\downarrow-1$. Evidently, any  such a contact gives a contribution $\epsilon$ to the system energy. 
Therefore, using  the definitions of $n^\uparrow$ and $n^\downarrow$ we can cast the  total energy of  the system  for a given 
microscopic configuration $\sigma$ as
\begin{equation}
\label{I}
H_\sigma=\epsilon\left[\sum_{k=1}^\infty (2-k)n^\uparrow_k+\sum_{k=1}^\infty (2-k) n^\downarrow_k-1\right].
\end{equation}
Obviously, the total number of spins $N_\sigma\equiv\sum_k(n^\uparrow_k+n^\downarrow_k)k$ should 
coincide with the lattice size $N$. Since up and down clusters alternate each other, then the only configurations 
with $n^\uparrow-n^\downarrow=0,~\pm1$ can be realized in the considered system. Hence, taking into account the
fact that the  total number of  cluster permutations is $n^\uparrow! n^\downarrow!/\prod_k n^\uparrow_k!n^\downarrow_k!$ 
we write the degeneracy factor of the microscopic configuration $\sigma$ as
\begin{eqnarray}
\label{II}
G_\sigma=\frac{\delta_{N,N_\sigma}
(2\delta_{n^\downarrow,n^\uparrow}+\delta_{n^\downarrow+1,n^\uparrow}+\delta_{n^\downarrow,1+n^\uparrow})
n^\uparrow! n^\downarrow!}{\prod\limits_{k=1}^\infty n^\uparrow_k!n^\downarrow_k!},
\end{eqnarray}
where $\delta$ is the Kronecker delta-symbol.  Note that  every configurations with $n^\downarrow=n^\uparrow$ should 
have a degeneracy 2, since it can be ``constructed'' only from the configurations $n^\downarrow+1=n^\uparrow$ and  
$n^\downarrow=1+n^\uparrow$ by combining one of their  outer clusters with any other cluster of the spin sign which 
coincides with the one of  outer cluster.

The $q^{th}$ statistical moment of the down cluster occupancy number we define as
\begin{equation}
\label{III}
\langle {n^\downarrow _l}^q\rangle=\frac{1}{Z}\sum_\sigma 
\left[n^\downarrow_l \right]^q G_\sigma \, e^{-\frac{H_\sigma}{T}+a\beta(N_\sigma-N)},
\end{equation}
where $T$ is the temperature and  $\beta=\epsilon/T$ is the lattice constant. A  summation over the configurations 
$\sigma$  in (\ref{III}) should be understood as a summation over all nonnegative integers $n_k^\uparrow$ and 
$n_k^\downarrow$. The dimensionless  parameter $a$ is introduced into  (\ref{III}) for the  convenience of analytical 
calculations and  its value is chosen to satisfy the conditions $\beta(a+1)<0$ and $\beta a <-\ln\left(2\cosh\beta\right)$. 
However,  this parameter does not affect any observable due to the condition $N_\sigma=N$ provided by the Kronecker 
delta-symbols in Eq.  (\ref{II}).   

Note, that $q=0$ yields the partition function $Z$, whereas a symmetry between up and down spins leads to the 
equality $\langle {n^\uparrow _l}^q\rangle=\langle {n^\downarrow _l}^q\rangle$. Using Eqs. (\ref{I}) and (\ref{II}),  
the definition of $N_\sigma$ along with the integral representation of the Kronecker delta-symbol 
$\delta_{r,s}=\int_0^{2\pi}\frac{d\alpha}{2\pi}e^{i\alpha(r-s)}$
and of  the factorials  $n^{\uparrow\downarrow}!=\int_0^\infty dz~e^{-z} z^{n^{\uparrow\downarrow}}$
we can write
\begin{eqnarray}
\label{IV}
\langle {n^\downarrow _l}^q\rangle&=&
\frac{e^{\beta(1-aN)}}{Z}
\int\limits_0^{2\pi}\frac{d\psi}{2\pi}e^{i\psi N} \int\limits_0^{2\pi}\frac{d\phi}{2\pi}(2+e^{i\phi}+e^{-i\phi})
\int\limits_0^\infty  dx~e^{-x}\int\limits_0^\infty  dy~e^{-y}\nonumber \\
& & \prod_{k=1}^\infty\sum_{n_k^\uparrow=0}^\infty
\frac{\left(xe^{\beta(k(a+1)-2)-i\phi -i\psi k}\right)^{n^\uparrow_k}}{n^\uparrow_k!}
\prod_{k=1}^\infty\sum_{n_k^\downarrow=0}^\infty
\frac{\left(ye^{\beta(k(a+1)-2)+i\phi -i\psi k}\right)^{n^\downarrow_k}}{n^\downarrow_k!}~{n^\downarrow_l}^q.~
\end{eqnarray}
Now the 
summations over $n^\uparrow_k$ and $n^\downarrow_{k\neq l}$  in (\ref{IV}) can be performed trivially giving  
the corresponding  exponential functions. In what follows we are mainly interested in finding the statistical moments 
of $n^\downarrow _l$ of  the  zeroth and first order. For   $q = \{0;  1\}$  the summation over $n^\downarrow_l$ generates 
an  exponential function  multiplied by the factor $(ye^{\beta(l(a+1)-2)+i\phi -i\psi l})^q$. Then,  each product over 
$k$ in (\ref{IV})  can be carried out,  since it  is  equivalent to a summation of  decreasing geometric  progressions in 
the exponential. In addition, the chosen range of $a$ provides a convergence of the integrals over $x$ and $y$  
variables due to negative real part of these exponential functions. Hence, for $q=0$ and $1$ one can easily find
\begin{eqnarray}
\label{V}
\langle {n^\downarrow _l}^q\rangle=\frac{e^{\beta(q(l-2)+1)}}{Z}\oint\limits_{|\zeta|=e^{-a\beta}}
\frac{d\zeta~\zeta^{N-ql-1} }{2\pi i}\oint\limits_{|\xi|=1}\frac{d\xi~(2+\xi+\xi^{-1}) \xi^q}
{2\pi i\left(\xi-\frac{e^{-\beta}}{\zeta-e^{\beta}}\right)\left(1-\frac{\xi e^{-\beta}}{\zeta-e^{\beta}}\right)^{q+1}},
\end{eqnarray}
where the real variables of integration $\phi$ and $\psi$ are now changed to the complex ones $\xi=e^{i\phi}$ and
$\zeta=e^{i\psi-a\beta}$. The integrand with respect to the variable $\xi$ in Eq. (\ref{V}) contains three poles. The 
first one $\xi=0$ exists for  $q=0$ only. However, the contribution of this pole does not produce any singularity with 
respect to the $\zeta$ variable and, hence, it can be safely neglected. The second  pole existing at any value of $q$ 
corresponds to $\xi=\frac{\zeta-e^{\beta}}{e^{-\beta}}$. It does not contribute to $\langle {n^\downarrow _l}^q\rangle$ 
since it is located out of the contour $|\xi|=1$ due to the chosen  value of the regularization parameter $a$. Thus, 
accounting only for  the simple pole $\xi=\frac{e^{-\beta}}{\zeta-e^{\beta}}$  we obtain
\begin{eqnarray}
\label{VI}
\langle {n^\downarrow _l}^q\rangle=
\frac{e^{\beta(q(l-3)+2)}}{Z}\oint\limits_{|\zeta|=e^{-a\beta}}\frac{d\zeta}{2\pi i }
\frac{\zeta^{N-ql-1} (\zeta-2\sinh\beta)^{1-q}(\zeta-e^{\beta})^{q+1}}{(\zeta-2\cosh\beta)^{q+1} }.
\end{eqnarray}
This expression clearly demonstrates that for $q=1$ and $l=N$  the  integral function has  a specific simple pole $\zeta=0$ 
which, as we discuss below,  is responsible for the non-monotonic behavior of $\langle n^\downarrow _l\rangle$.

Taking $q=0$ we immediately  reproduce the well-known partition function of the one dimensional Ising model with free edges \cite{Ising}, i.e.
\begin{eqnarray}
\label{VII}
Z=e^{2\beta}\oint\limits_{|\zeta|=e^{-a\beta}}
\frac{d\zeta~\zeta^{N-1} (\zeta-2\sinh\beta)(\zeta-e^{\beta})}{2\pi i(\zeta-2\cosh\beta) }=2(2\cosh\beta)^{N-1}.
\end{eqnarray}
For $l\le N$ the average occupancy number of the down clusters is obtained from Eq. (\ref{VI}) for $q=1$
\begin{eqnarray}
\label{VIII}
\langle n^\downarrow _{l\le N}\rangle&=&\frac{e^{\beta(1-l)}}{Z}\oint\limits_{|\zeta|=
e^{-a\beta}}\frac{d\zeta~\zeta^{N-l-1} (\zeta-e^{\beta})^2}{2\pi i(\zeta-2\cosh\beta)^2}
=\frac{\delta_{N,l}+2e^{-2\beta}+(N-l+1)e^{-4\beta}}{2(1+e^{-2\beta})^{l+1}},
\end{eqnarray}
whereas $\langle n^\downarrow _{l>N}\rangle=0$. Note, that $\langle n^\uparrow _l\rangle=\langle n^\downarrow _l\rangle$
for all $l$. The direct calculation based on this expression demonstrates that 
$\sum_l l \langle n^\uparrow _l\rangle=\sum_l l \langle n^\downarrow _l\rangle=N/2$ which is an obvious consequence of the symmetry 
between the spins up and down,  if  the external magnetic field is absent.

An explicit expression (\ref{VIII}) for $\langle n^\downarrow _l\rangle$ allows us to find the  size distribution function of clusters 
$\omega_l$ which is proportional to their average occupancy number. Since the size distribution functions are the same for both 
kinds of clusters, we do distinguish them. The  normalized  distribution $\omega_l$ is obtained from the condition that the total 
probability to find up (or down) cluster of any size is  $1/2$,  i. e. $\sum_l \omega_l=1/2$. This condition follows from the symmetry 
between up and down spins. Hence, we obtain
\begin{eqnarray}
\label{IX}
\omega_{l\le N}=\frac{\delta_{N,l}+2e^{-2\beta}+(N-l+1)e^{-4\beta}}{2(1+Ne^{-2\beta})(1+e^{-2\beta})^l}\,,
\end{eqnarray}
and $\omega_{l> N}=0$.  From Eq. (\ref{IX}) one can see that $\omega_l$ has not only the exponential part  $(1+e^{-2\beta})^{-l}$, but  
also it includes the factor which is  linear in $l$. The power part of the size distribution functions is known in several statistical  models of  cluster 
type. Traditionally it  is taken into account  by the Fisher topological exponent $\tau$ as $l^{-\tau}$ \cite{Fisher,Bondorf,CSMM,QGBSTM}. 
However, the linear $l$-part of the size distribution function (\ref{IX}) disappears in the thermodynamic limit $N\rightarrow\infty$, since in this 
limit one finds $\omega_l\rightarrow e^{-2\beta}(1+e^{-2\beta})^{-l}/2$. Hence, it is  interesting to analyze the distribution  $\omega_l$  of  
other lattice models in order to  find possible restrictions on values of the Fisher index $\tau$.


\section{Bimodality manifestation}
\label{Bimodality}

The growing interest to a bimodality is caused by the widespread belief that it can serve as a robust signal of the  first order phase transition in finite  systems.  As it was mention above this concept is based on T. L. Hill idea \cite{THill:1} that, each peak  of the bimodal distribution is associated with a pure phase. However, this idea cannot be sufficiently justified,  since in finite systems  the analog of  mixed  phase  is not just a mixture of two pure phases \cite{Bugaev05,CSMM,Bmodal:Bugaev13}.  According to exact analytical solutions found for  several cluster models in finite systems \cite{Bugaev05,CSMM,Bmodal:Bugaev13}  the finite volume analog of  mixed  phase is  represented by a combination of  an analog of gaseous phase, which is stable, and some even number of different  metastable 
states. The number of metastable states is determined by thermodynamic parameters, of course, but the whole point is that their  identification  with the help of  a single peak (or maximum) of the size distribution function 
does not  seem to be a well defined procedure \cite{Bugaev05,CSMM,Bmodal:Bugaev13}.

Having at hand  an explicit expression for cluster size distribution we are able to answer the question what is the reason for a bimodal behavior. 
 From Eq. (\ref{IX}) it follows  that  $\omega_l=\delta_{N,l}/2$ for  finite $N$ and for  $\beta\rightarrow\infty$ (a 
 ferromagnetic system at zero 
temperature), whereas  the limit $\beta\rightarrow-\infty$ (an antiferromagnetic system at zero temperature) generates $\omega_l=\delta_{1,l}/2$ 
for  finite $N$. Therefore, we conclude that at some intermediate value of the lattice constant $\beta$ the monomer dominant regime should switch
to the regime of dominance of  the cluster of maximal size. This switching is characterized by the comparable probabilities to find the monomers (dimers, 
trimers, etc.) and the cluster of maximal size and, hence, at some value of $\beta$ the distribution function $\omega_l$ becomes a non-monotonic one. 

A direct inspection of  Eq. (\ref{IX}) shows that for $l< N$ and finite $N$ the derivative $\frac{\partial \omega_l}{\partial l}$ 
is always negative, i.e. $\frac{\partial \omega_l}{\partial l} < 0 $ and, hence, at this interval of cluster sizes the function 
$\omega_l$ is a monotonic one. The  non-monotonic behavior of the distribution $\omega_l$ is caused by presence of the 
Kronecker delta-symbol  $\delta_{N,l}$. It is clear that the cluster size distribution is non-monotonic, if  the inequality 
$\omega_{N-1}<\omega_N$ is obeyed or, equivalently, if  
\begin{eqnarray}\label{EqX}
\beta>\beta_c\equiv\ln2/2 \,.
\end{eqnarray}
This  inequality can be fulfilled for $\epsilon>0$, only.  This  means that the non-monotonicity of $\omega_l$ 
in the one dimensional  Ising model can appear in the ferromagnetic  case only. It  is necessary  to stress that the value of $\beta_c$  does not depend on $N$ and $\epsilon$ and, hence, it is a universal constant. 
Thus, in the ferromagnetic case the found cluster distribution function $\omega_l$ is non-monotonic  (bimodal)
for any size of the lattice and any value of the spin coupling.
However, we should note that the bimodal behavior of the cluster size 
distribution is washed out in the thermodynamic limit $N\rightarrow\infty$, since in this case
$\omega_N-\omega_{N-1}\rightarrow0$ for all  finite values of the lattice constant $\beta$.

The mathematical reason of  the non-monotonic behavior of cluster size distribution  in the present model is clear now. 
Although in this model the  bimodality phenomenon appears as the finite size effect, we would like 
to add a few words on its  substantial origin.
For this purpose let us  
consider an auxiliary infinite lattice with the cluster distribution function $\tilde\omega_l$ which for all $l$ obeying inequality 
$N < l < l_0$ is defined as
\begin{eqnarray}
\label{X}
\tilde\omega_l=\frac{2e^{-2\beta}+(N-l+1)e^{-4\beta}}{2(1+Ne^{-2\beta})(1+e^{-2\beta})^l} \,,
\end{eqnarray}
where the parameter $l_0 = N+1 + 2\,e^{2\beta}$ provides the nonnegative value of $\tilde\omega_{l_0}$.
For $\beta \gg 1$
the distribution (\ref{X}) describes clusters of almost all sizes (up to $l_0 \gg N$) on an equal footing. 
 Note, that  in Eq. (\ref{X}) the quantity $N$  is treated as a finite parameter. 
 Suppose now that we randomly choose on this infinite lattice the infinite amount of  intervals of the length $N$ each.
 It is clear that the size distribution of small clusters will be given by  Eq. (\ref{X}), whereas the clusters with $l > N$ will 
increase the occupancy number   of  the cluster of maximal size (they will also contribute to the smaller clusters, but this fact is already taken into account in  (\ref{IX})). This can be easily seen, if one sums up all values of $\tilde\omega_l$ for $l > N$
\begin{eqnarray}
\label{XI}
\sum_{l=N+1}^\infty\tilde\omega_l=\frac{1}{2(1+Ne^{-2\beta})(1+e^{-2\beta})^N}=\omega_N-\tilde\omega_N \,.
\end{eqnarray}
Although   in the sum above the terms  with $l > l_0$ are negative, their contribution is practically negligible due to the adopted assumption that $\beta \gg 1$.  The result of summation is nothing else, but  the term in  $\omega_N$ of  Eq. (\ref{IX}) which is proportional to the Kronecker delta-symbol. 
Eq. (\ref{XI}) demonstrates   that in a finite system the clusters of size larger than the  lattice size are ``condensing''  into the largest cluster  of size $N$, whereas their  probabilities $\tilde\omega_{N<l<l_0}$ are ``absorbed''  into $\omega_N$.

\begin{figure}[h]
\centerline{\includegraphics[width=0.60\columnwidth]{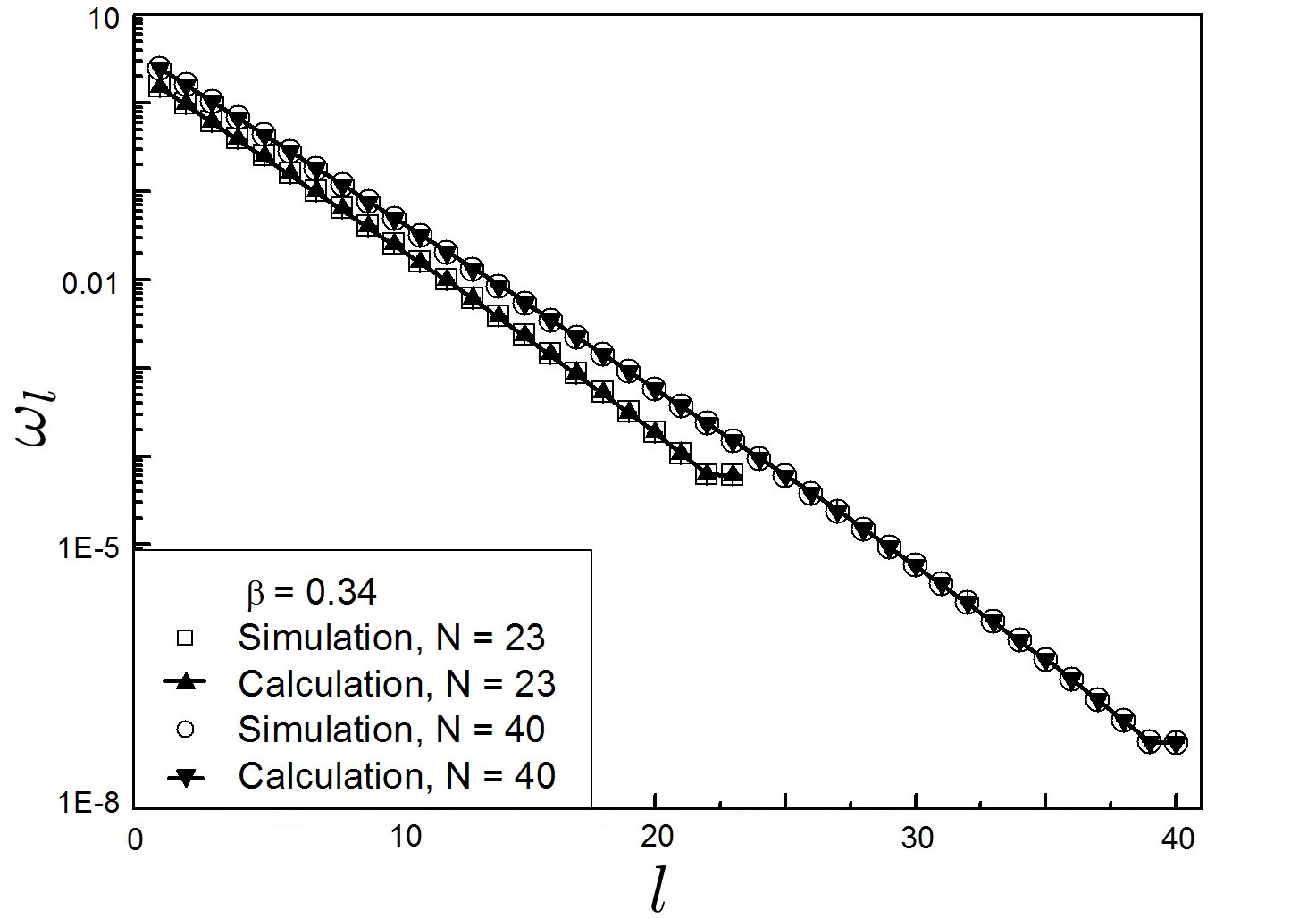}}
\centerline{\includegraphics[width=0.60\columnwidth]{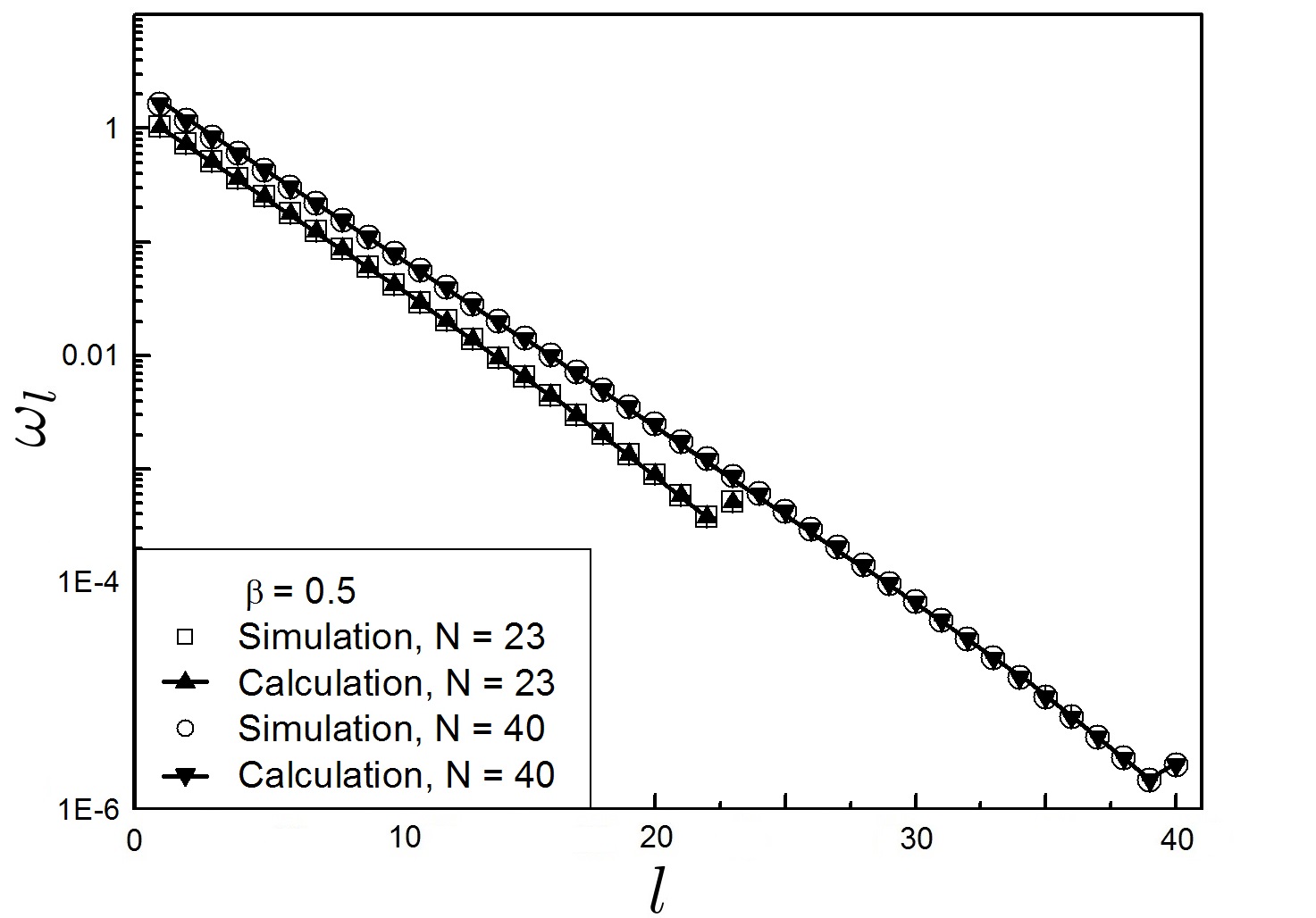}}
\centerline{\includegraphics[width=0.60\columnwidth]{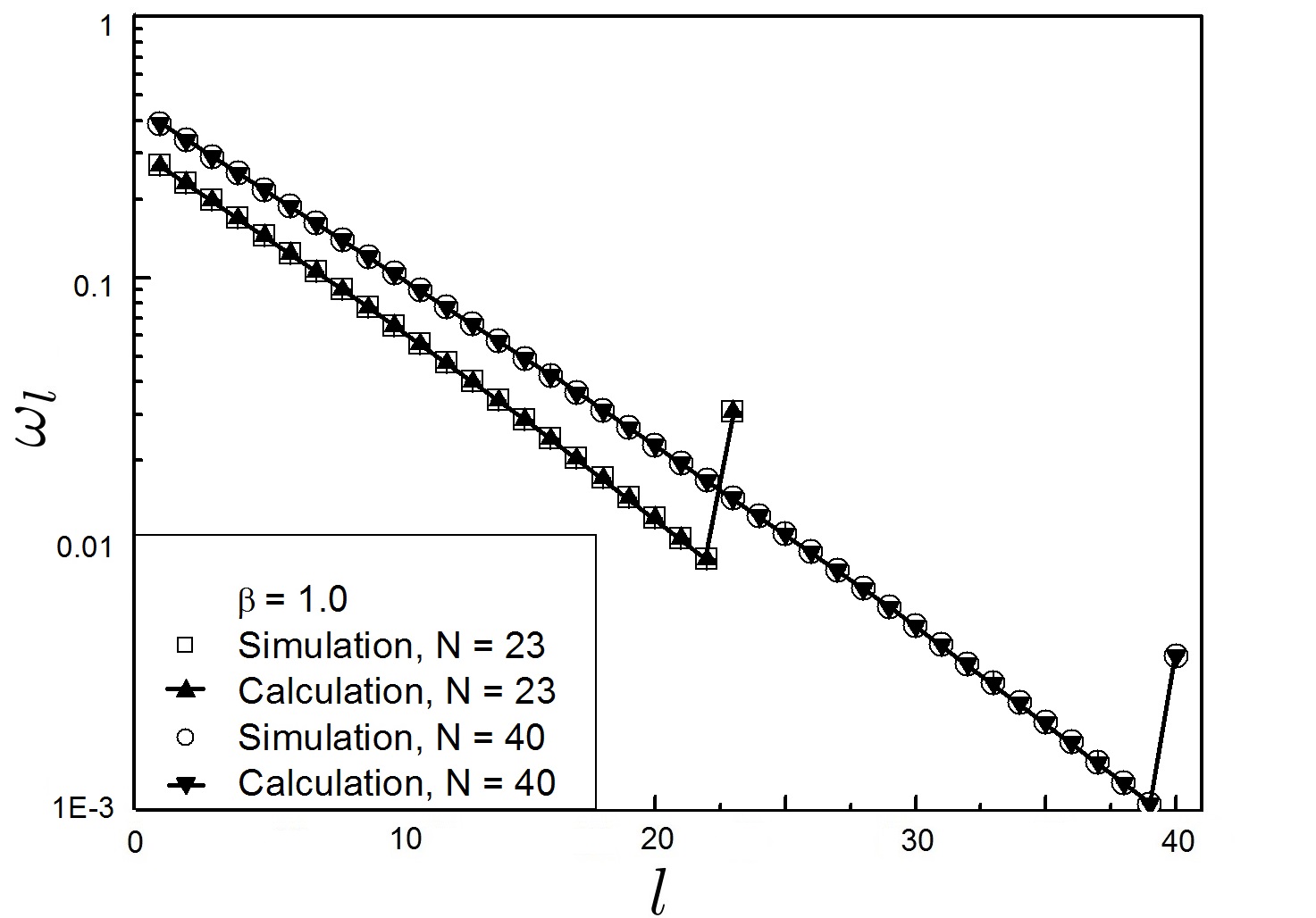}}
\caption{Cluster size distribution calculated using  Eq. (\ref{IX}) (triangles) and found  numerically  for $N=23$ and 
$N=40$ (squares and circles) for three values of the lattice constant $\beta$. 
For  $\beta=0.34 < \beta_c$ (upper panel) $\omega_l$ monotonically 
decreases with the cluster size growth, while for $\beta=0.5 > \beta_c$ (middle panel) and $\beta=1.0 > \beta_c$ (lower panel) it has a maximum at $l=N$.}
\label{fig}
\end{figure}

In order to demonstrate the occurrence of  bimodality in the one dimensional Ising model we present here the results of our numerical 
simulations for the lattice sizes $N=23$ and $N=40$. 
These simulations were made using the Swendsen-Wang algorithm. 
For each run the first $5\cdot 10^5$   lattice configurations were discarded to ensure a complete thermalization, and the next $5 \cdot 10^7$ lattice measurements were made while  discarding every 5 lattice configurations between the measurements. For each measurement the total 
expectation of the number of clusters of  all  sizes  was calculated, then  the jackknife error analysis was used to obtain the error estimate 
of the mean values obtained. 

Our numerical study includes 14 values of the lattice constant in the range from 0 to 2.5. We would like to stress  that the
coincidence of distribution functions simulated numerically and calculated according to formula ($\ref{IX}$) 
is  perfect. In Fig. \ref{fig} the cluster size distribution is shown for three values of the lattice constant $\beta$. For
$\beta=0.34<\beta_c$ the cluster size distribution is monotonic for both values of $N$. At the same time for the
values of lattice constant  above the ``critical'' one, i.e. for $\beta=0.5  >\beta_c$ and for $\beta=1.0  >\beta_c$,  one can 
clearly see the bimodal distribution in Fig. \ref{fig}.
It is also seen that bimodality is enhanced with the lattice constant increasing. 
We have to remind once again  that  the present model
does not have any phase transition  in the thermodynamic limit. Hence, the one dimensional Ising model gives
an explicit counterexample to treating bimodality as a signal of phase transition in finite systems. 


\section{Conclusions and perspectives} 
\label{Conclusions}

In this work we found an analytical solution of the one dimensional Ising model in terms of  the geometrical clusters
composed of the neighboring spins of the same direction. It is clear that the present  formulation  is an important step 
towards 
 establishing  a firm  connection  between the lattice spin systems  and the statistical  models of  cluster type.
The developed  approach allows  us to exactly calculate   the size distribution of spin clusters for  finite and infinite lattice sizes. 
Using the exact formulae we showed that  for finite size of the lattice   the one dimensional Ising model 
has a bimodal size distribution of clusters  for $\beta > \ln 2/2$. 
We demonstrate  that  the bimodal size
distribution of spin clusters  in the present model is a finite size effect which  appears due to a ``condensation'' of 
clusters whose  size exceeds  the lattice extent  to the largest  cluster. 
Such a  ``condensation''  to the maximal cluster 
leads to  the non-monotonic size distribution at sufficiently large values of the lattice constant in ferromagnetic case only. 
The obtained  result is valid for all sizes of the lattice, however, in the thermodynamic limit it is washed out. 
It provides us with  an explicit counterexample to the widely spread  belief about an exclusive role of  bimodality as a signal of phase transition in finite systems \cite{THill:1,Bmodal:Chomaz01,Bmodal:Chomaz03,Bmodal:Gulm04,Bmodal:Gulm07}. Our  criticism is in full coherence with the results of  Refs. \cite{CSMM,Lopez,Bmodal:Bugaev13}.
The developed  formalism considers  the simplest exactly solvable lattice model, but it would be interesting to apply it 
to more realistic physical systems. 


\mbox{}\\
\hfill\\
\noindent
{\bf Acknowledgments.}
The authors appreciate the fruitful discussions with K. A. Bugaev and  V. K. Petrov and their valuable comments. 
The work of  A. I. I.  is supported in part by   the National Academy of Sciences of Ukraine Grant of GRID simulations for high energy physics. 



\begin{thebibliography}{99}

\bibitem{Gross}
D. H. E. Gross, Phys. Rep. 279, 119 (1997).

\bibitem{Bmodal:Chomaz01}
Ph. Chomaz, F. Gulminelli and V. Duflot,  Phys. Rev. E, 64, 046114 (2001).

\bibitem{ComplMoretto}
 L. G. Moretto et al., Phys. Rev. Lett., 94, 202701 (2005).

\bibitem{CSMM}
K. A. Bugaev, Phys. Part. Nucl., 38, 447 (2007).

\bibitem{JoeNat}
J. Natowitz et al., Phys. Rev. C, 65, 034618 (2002).

\bibitem{Pichon}
Pichon M. et al. (INDRA and ALADIN Collaborations), Nucl. Phys. A, 779, 267 (2006).

\bibitem{Bonnet}
E. Bonnet  et al. (INDRA and ALADIN Collaborations), Phys. Rev. Lett., 103, 072701 (2009).

\bibitem{Pratt}
S. Pratt, Physics 1, 29 (2008)

\bibitem{Karsch13}
F. Karsch, PoSCPOD, 2013, 046 (046).

\bibitem{Chomaz03}
F. Gulminelli, Ph. Chomaz, Al. H. Raduta, and Ad. R. Raduta, Phys. Rev. E, 64, 046114 (2001).

\bibitem{Bugaev05}
Bugaev  K. A., Acta. Phys. Polon., B 36, 3083 (2005).



\bibitem{Bmodal:Bugaev13}
K. A. Bugaev,  A. I. Ivanytskyi, V. V. Sagun, D. R. Oliinychenko, Phys. Part. Nucl. Lett., 10, 832 (2013).


\bibitem{THill:1}
T. L. Hill, Thermodynamics of small  systems, Dover, New York (1994).

\bibitem{Bmodal:Chomaz03}
Ph. Chomaz and F. Gulminelli, Physica A, 330, 451 (2003).

\bibitem{Bmodal:Gulm04}
F. Gulminelli, Ann. Phys. Fr., 29, 6 (2004).

\bibitem{Bmodal:Gulm07}
F. Gulminelli, Nucl. Phys. A, 791, 165 (2007).

\bibitem{YangLee}
C. N. Yang  and  T. D. Lee, Phys. Rev., 87, 404 (1952).

\bibitem{Lopez}
O. Lopez,  D. Lacroix, and E. Vient, Phys. Rev. Lett., 95, 242701 (2005).

\bibitem{Mekjian}
S. Das Gupta and A. Z. Mekjian, Phys. Rev. C, 57, 1361 (1998).

\bibitem{Sagun14}
V. V. Sagun, A. I. Ivanytskyi, K. A. Bugaev and I. N. Mishustin, Nucl. Phys. A, 924, 4, 24  (2014).

\bibitem{Ising}
E. Ising, Z. Phys., 31, 253 (1925).

\bibitem{Fisher}
M. E. Fisher, Physics, 3, 255 (1967).

\bibitem{Dillmann}
A. Dillmann and G. E. Meier, J. Chem. Phys. 94, 3872 (1991).

\bibitem{Ford}
A. Laaksonen, I. J. Ford, and M. Kulmala, Phys. Rev. E 49, 5517 (1994).

\bibitem{Bondorf}
J. P. Bondorf et al., Phys. Rep., 257, 131 (1995).

\bibitem{Kapusta}
J. I. Kapusta J. I., Phys. Rev.  D, 23, 2444 (1981).

\bibitem{Gorenstein}
M. I. Gorenstein, G. M. Zinovjev and V. K. Petrov, Phys. Lett. B, 106, 327 (1981).

\bibitem{QGBSTM}
K. A. Bugaev, Phys. Rev. C, 76, 014903 (2007).

\bibitem{QGBSTM2}
K. A. Bugaev, V. K. Petrov and G. M. Zinovjev,  Phys.  Part. Nucl. Lett., 9, 3, 238  (2012). 

\bibitem{QGBSTM3}
K. A. Bugaev, A. I. Ivanytskyi, E. G. Nikonov, V. K. Petrov,  A. S. Sorin and G. M. Zinovjev,
Phys. Atom. Nucl., 75, 6, 707 (2012). 

\bibitem{CGreiner:06}
I. Zakout, C. Greiner, J. Schaffner-Bielich, Nucl. Phys., A 781,  150 (2007).

\bibitem{CGreiner:07}
I. Zakout, C. Greiner,  Phys. Rev. C, 78, 034916 (2008). 

\bibitem{Koch:09}
L. Ferroni and V. Koch, Phys. Rev. C, 79, 034905 (2009). 


\bibitem{Moretto}
L. G. Moretto et al., Phys. Rev. C, 68,1602 (2003) and references therein.

\bibitem{Fortunato2000}
S. Fortunato and H. Satz, Phys. Lett. B, 475, 311 (2000).

\bibitem{Fortunato2001}
S. Fortunato et. al, Phys. Lett. B, 502, 321 (2001).

\bibitem{Gattringer2010}
C. Gattringer, Phys. Lett. B, 690, 179 (2010).

\bibitem{Gattringer2011}
C. Gattringer and A. Schmidt, JHEP, 1101, 051 (2011.


\bibitem{Regensburg15}
G. Endrodi,  A. Sch{\"a}fer and  J. Wellnhofer, arXiv:1506.07698 [hep-lat]. 

\bibitem{Kiev15}
A. I. Ivanytskyi et. al., Physical properties of  Polyakov loop geometrical clusters in  $SU(2)$ gluodynamics (2015) (in preparation).


\end{thebibliography}
\end{document}